\documentclass[journal=nalefd,manuscript=letter]{achemso}

\author{V\'ictor Fern\'andez-Hurtado}
\affiliation{Departamento de F\'{\i}sica Te\'orica de la Materia Condensada and
Condensed Matter Physics Center (IFIMAC), Universidad Aut\'onoma de Madrid,
E-28049 Madrid, Spain}
\alsoaffiliation{Department of Physics, University of Konstanz, D-78457 Konstanz, Germany}
\email{victor.fernandezh@uam.es}
\author{Antonio I. Fern\'andez-Dom\'inguez}
\affiliation{Departamento de F\'{\i}sica Te\'orica de la Materia Condensada and
Condensed Matter Physics Center (IFIMAC), Universidad Aut\'onoma de Madrid,
E-28049 Madrid, Spain}
\author{Johannes Feist}
\affiliation{Departamento de F\'{\i}sica Te\'orica de la Materia Condensada and
Condensed Matter Physics Center (IFIMAC), Universidad Aut\'onoma de Madrid,
E-28049 Madrid, Spain}
\author{Francisco J. Garc\'ia-Vidal}
\affiliation{Departamento de F\'{\i}sica Te\'orica de la Materia Condensada and
Condensed Matter Physics Center (IFIMAC), Universidad Aut\'onoma de Madrid,
E-28049 Madrid, Spain}
\alsoaffiliation{Donostia International Physics Center (DIPC), Donostia/San Sebasti\'an 20018, Spain}
\author{Juan Carlos Cuevas}
\affiliation{Departamento de F\'{\i}sica Te\'orica de la Materia Condensada and
Condensed Matter Physics Center (IFIMAC), Universidad Aut\'onoma de Madrid,
E-28049 Madrid, Spain}
\affiliation{Department of Physics, University of Konstanz, D-78457 Konstanz, Germany}
\alsoaffiliation{Department of Physics, University of Konstanz, D-78457 Konstanz, Germany}

\title{Exploring the limits of super-Planckian far-field radiative heat transfer using 2D materials}

\begin{document}

%%%%%%%%%%%%%%%%%%%%%%%%%%%%%%%%%%%%%%%%%%%%%%%%%%%%%%%%%%%%%%%%%%%%%
\begin{tocentry}

\centering\includegraphics[width=6.8cm]{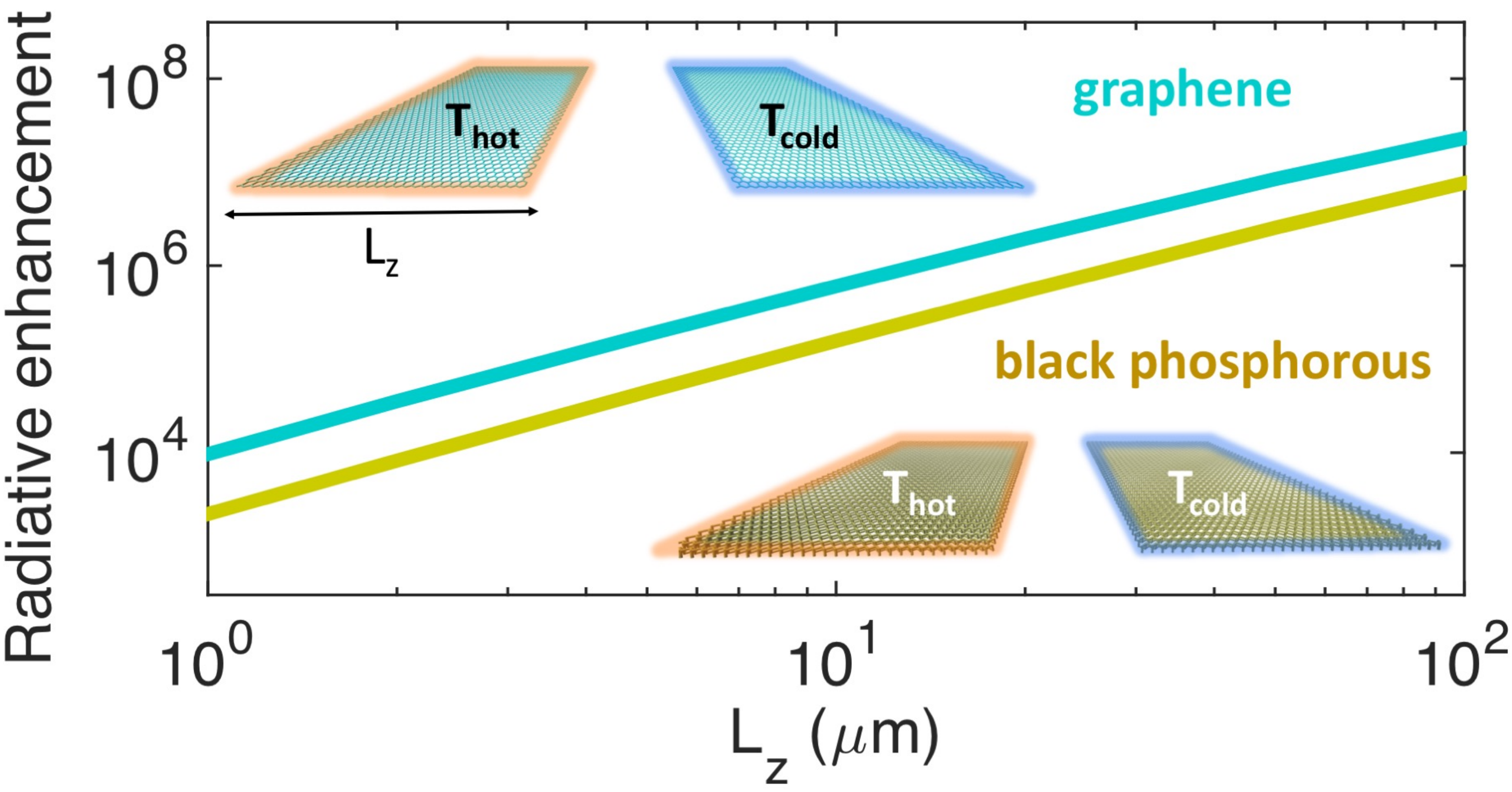}

\end{tocentry}

%%%%%%%%%%%%%%%%%%%%%%%%%%%%%%%%%%%%%%%%%%%%%%%%%%%%%%%%%%%%%%%%%%%%%
\begin{abstract}
Very recently it has been predicted that the far-field radiative heat transfer between two 
macroscopic systems can largely overcome the limit set by Planck's law if one of their dimensions 
becomes much smaller than the thermal wavelength ($\lambda_{\rm Th} \approx 10\, \mu$m at room 
temperature). To explore the ultimate limit of the far-field violation of Planck's law, here we present a
theoretical study of the radiative heat transfer between two-dimensional (2D) materials.
We show that the far-field thermal radiation exchanged by two coplanar systems with a one-atom-thick 
geometrical cross section can be more than 7 orders of magnitude larger than the theoretical
limit set by Planck's law for blackbodies and can be comparable to the heat transfer of two parallel
sheets at the same distance. In particular, we illustrate this phenomenon with 
different materials such as graphene, where the radiation can also be tuned by a external gate,
and single-layer black phosphorus. In both cases the far-field radiative heat transfer is dominated by 
TE-polarized guiding modes and surface plasmons play no role. Our predictions provide a 
new insight into the thermal radiation exchange mechanisms between 2D materials.
\end{abstract}
%%%%%%%%%%%%%%%%%%%%%%%%%%%%%%%%%%%%%%%%%%%%%%%%%%%%%%%%%%%%%%%%%%%%%

Radiation is, together with convection and conduction, one of the three basic mechanisms of 
heat exchange between bodies \cite{Incropera2017}. The maximum thermal energy that can be 
transferred between two objects via radiation is, in principle, set by Planck's law for 
blackbodies \cite{Planck1914}, which assumes that both of them are perfect absorbers at all 
frequencies and that all dimensions involved in the problem are larger than $\lambda_{\rm Th}$. 
However, it is known that when the separation between two bodies is smaller than 
$\lambda_{\rm Th}$, the radiative heat transfer can be enhanced by orders of magnitude due to 
the contribution of evanescent waves \cite{Polder1971,Rytov1989,Joulain2005,Basu2009,Song2015}. 
This phenomenon, known as near-field radiative heat transfer (NFRHT) \cite{Polder1971,Rytov1989}, 
has been confirmed in recent years by several experiments exploring different geometries, materials 
and distances between the two objects, ranging from micrometers down to a few nanometers \cite{Kittel2005,
Rousseau2009,Shen2009,Ottens2011,Kralik2012,Zwol2012a,Worbes2013,St-Gelais2014,Song2015a,Kim2015,
St-Gelais2016,Song2016,Bernardi2016,Cui2017,Kittel2017}. Very recently, we have predicted that 
the Planckian limit can also be largely surpassed in the far-field regime \cite{Fernandez2017}, i.e.,
when the separation of the objects is larger than $\lambda_{\rm Th}$. In particular, we have 
shown theoretically that the far-field radiative heat transfer (FFRHT) between micron-size devices
can overcome the black-body limit by several orders of magnitude if their thickness is much smaller 
than $\lambda_{\rm Th}$. Moreover, we have shown that the enhancement over Planck's law increases monotonically
as the device thickness is reduced \cite{Fernandez2017}, which leads us to the fundamental question 
on the limits of super-Planckian FFRHT. The goal of this work is to explore this issue with the help of 
2D materials, i.e., with materials with a one-atom-thick geometrical cross section, which constitute 
the ultimate limit of thin systems.

2D materials have been extensively studied in recent years in the context of radiative heat transfer. 
In particular, several works have taken advantage of the near-field density of photonic states in these 
systems to modify the characteristics of emitters in a wide variety of scenarios. Most of the 
theoretical work in the case of graphene has focused on the possibility to tune and enhance the NFRHT 
mediated by the surface plasmon-polaritons sustained by this material \cite{Volokitin2011,Ilic2012}. For 
instance, it has been predicted that the NFRHT between polar dielectrics can be boosted by placing a 
graphene layer on top \cite{Stetovoy2012,Liu2014,Messina2017}. This prediction has been confirmed 
experimentally \cite{Zwol2012b}. Other studies have proposed periodic graphene ribbon arrays to induce 
hyperbolic modes and thus further enhance the NFRHT between 2D systems \cite{Liu2015}. The NFRHT between 
graphene nanodisks has also been studied \cite{Ramirez2017}, and the analysis of the time scales of radiative 
heat transfer in this setup suggests that this process is ultrafast \cite{Renwen2017}. Let us also mention 
that the near field thermal conductance between Dirac 2D materials scales as the inverse of the distance 
between two flakes \cite{Rodriguez2015}. However, and despite all these recent advances, FFRHT between 2D 
materials remains unexplored. As explained above, 2D materials constitute ideal systems in which one can 
explore the ultimate limit of the violation of Planck's law in the far-field regime. Moreover, from an 
applied viewpoint, understanding the absorption and emission of radiation in 2D 
materials is key to properly characterize their thermal properties and harness their
unique mechanical and electronic features \cite{Bastelle2011}. For these reasons, we present 
in this work a theoretical study of the FFRHT between systems with a one-atom-thick geometrical cross section. 
In particular, we demonstrate that the FFRHT between sheets of 2D materials like graphene or single-layer 
black phosphorus can overcome the Planckian limit by more than 7 orders of magnitude. Moreover, we show that, 
contrary to the known NFRHT mechanism, surface plasmon-polaritons play no role in this case and this 
remarkable heat transfer is instead dominated by TE-polarized guiding modes.

\begin{figure}[t]
\centering\includegraphics[width=12.3cm]{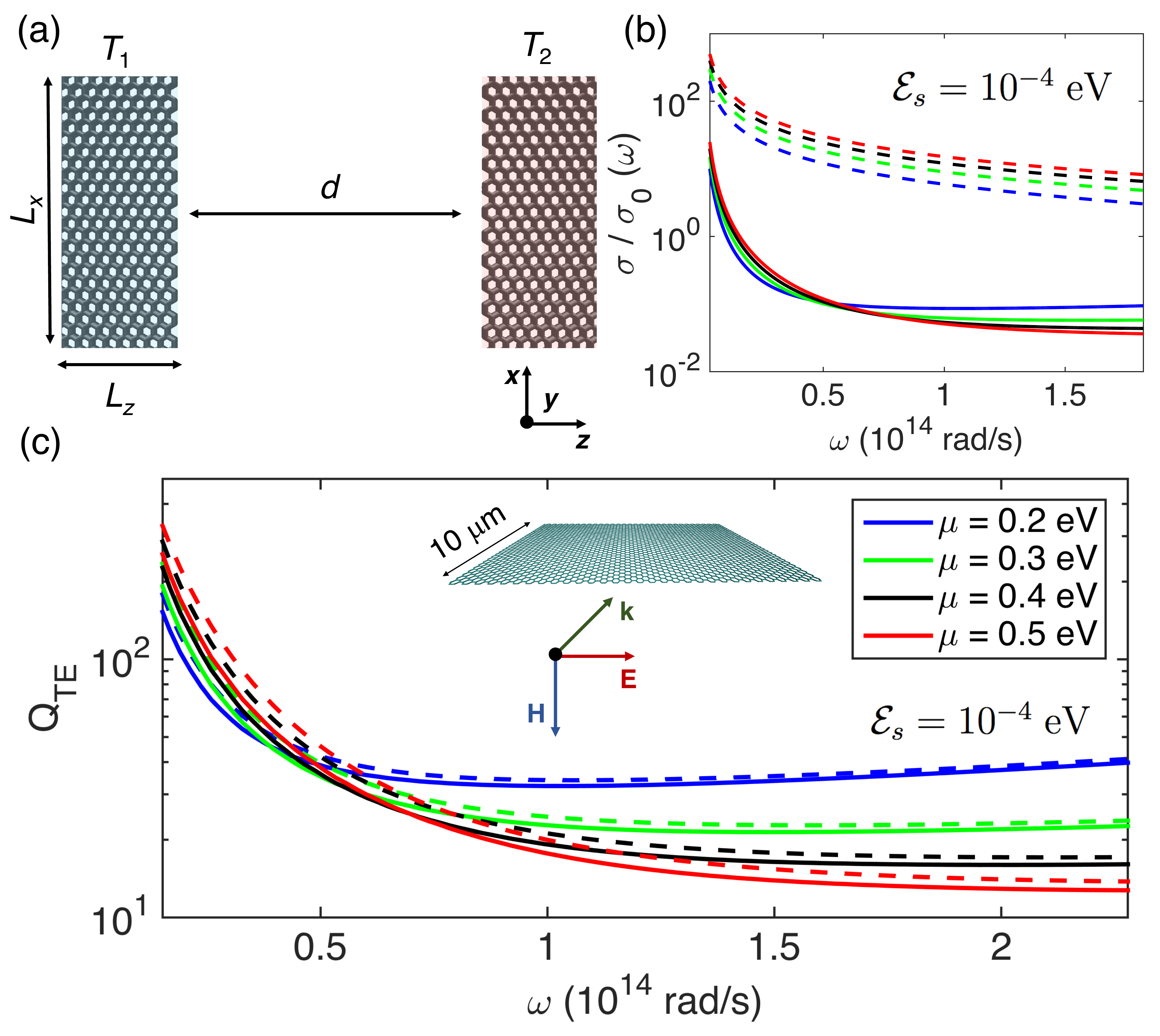}
\caption{(a) Schematics of the FFRHT between two identical graphene flakes. The flakes have dimensions $L_x \times L_z$, 
are separated by a gap $d$, and are held at temperatures $T_1$ and $T_2$, respectively. (b) Real (solid lines) and 
imaginary (dashed lines) part of the normalized conductivity for $T=300$ K, $\mathcal{E} _{\rm s}=10^{-4}$ eV, and for 
different chemical potentials ($\mu$), as indicated in the legend of panel (c). (c) Frequency-dependent absorption 
efficiency for a plane wave with transverse electric polarization ($Q_{\rm TE}(\omega)$) and normal incidence into a 
graphene sheet with length $L_z=10$ $\mu$m, infinite width (see inset), and for different values of $\mu$ (see legend). 
The solid lines correspond to the exact numerical results, while the dashed lines correspond to the results obtained
with eq~\ref{eq-anal}.}
\label{fig-setup}
\end{figure}

Let us start by analyzing the FFRHT between two coplanar graphene sheets. This system is schematically represented
in Figure~\ref{fig-setup}a. In this case, two identical graphene sheets at temperatures $T_{1}$ and  $T_{2}$ 
($T_{1}<T_{2}$) are separated by a gap $d$. The length of the flakes is denoted by $L_{z}$ and, for simplicity,
we shall assume that they are infinitely wide in the $x$-direction ($L_x \to \infty$). Notice that, as shown 
in Figure~\ref{fig-setup}a, both flakes are coplanar and thus, for the radiative problem they constitute systems with a
one-atom-thick geometrical cross section. In order to compute the power exchanged in the form of thermal radiation between 
these 2D systems, we make use of the theory of fluctuational electrodynamics \cite{Rytov1989,Joulain2005}.
In this theory, the material properties enter via the dielectric function, which in the graphene
case can be determined from the electrical conductivity\cite{Nikitin2013}. The 2D conductivity of 
graphene, $\sigma^{\rm graphene}_{\rm 2D}$, calculated within the random phase approximation can be expressed
 in terms of the chemical potential ($\mu$), temperature ($T$) and scattering energy 
($\mathcal{E} _{\rm s}$) \cite{Nikitin2013}
\begin{equation}
\sigma^{\rm graphene}_{\rm 2D} = \sigma_{\rm intra} + \sigma_{\rm inter} ,
\end{equation}
where the intraband and the interband contributions are given by
\begin{eqnarray}
\sigma_{\rm intra} &  =  & \frac{2ie^2t}{\hbar \pi (\Omega + i\gamma)} 
\ln \left[ 2 \cosh \left( \frac{1}{2t} \right) \right] , \nonumber \\
\sigma_{\rm inter} & = &\frac{e^2}{4 \hbar} \left[ \frac{1}{2} + \frac{1}{\pi} \arctan \left( \frac{\Omega-2}{2t} \right)
-\frac{i}{2\pi} \ln \frac{\left( \Omega+2 \right)^2}{\left( \Omega-2 \right)^2+\left(2t \right)^2} \right],
\end{eqnarray}
with $\Omega=\hbar \omega/\mu$, $\gamma=\mathcal{E} _{\rm s}/\mu$, and $t=k_{\rm B}T/\mu$. In 
Figure~\ref{fig-setup}b we show the normalized 2D conductivity of graphene, in units of $\sigma_0=e^2/2\pi \hbar$,
for $T=300$ K, $\mathcal{E}_{\rm s}=10^{-4}$ eV, and different values of the chemical potential. As one can
see, graphene resembles a Drude metal in the infrared regime whose metallic character increases with 
the chemical potential. Let us remark that the value chosen in this case for $\mathcal{E} _{\rm s}$ corresponds 
to a graphene sheet with a very large relaxation time, $\tau = \hbar/\mathcal{E}_{\rm s}$, which is normally the 
desired scenario in the field of graphene plasmonics \cite{Koppens2017}. We will show below that the opposite
limit is indeed more favorable for the absorption and emission of radiation between 2D materials.

In order to calculate the FFRHT, we make use of a result derived in a recent 
work\cite{Fernandez2017} with the help of a thermal discrete dipole approximation \cite{Martin2017}. This 
result establishes a connection between the FFRHT between two objects and their absorption efficiencies,
i.e., their absorption cross sections divided by their geometrical cross sections. Assuming that a sheet of a
2D material can be modeled as a parallelepiped (see below), this result indicates that the radiative power exchanged
between two identical flakes at temperatures $T_{1}$ and $T_{2}$ and separated by a gap $d$ much larger than both
$\lambda_{\rm Th}$ and their characteristic dimensions is given by \cite{Fernandez2017}
\begin{equation}
P = \frac{\pi}{2}  A F_{12} \int^{\infty}_0 [Q^2_{\rm TE}(\omega)+Q^2_{\rm TM}(\omega)] \left[ I_{\rm BB}(\omega,T_1) -
I_{\rm BB}(\omega,T_2) \right] d\omega ,
\label{eq-Q}
\end{equation}
where $A$ is the geometrical cross section of the bodies and $F_{12}=\delta/2d$ is the geometrical view factor \cite{Incropera2017},
where $\delta$ is the geometric thickness of the 2D material. On the other 
hand, $Q_{\rm TM, TE}(\omega)$ is the frequency-dependent absorption efficiency for a plane wave with normal incidence and 
transverse magnetic (TM) or transverse electric (TE) polarization, and $I_{\rm BB}(\omega,T)$ is the Planck distribution 
function, which is given by
\begin{equation}
I_{\rm BB}(\omega,T) = \frac{\omega^2}{4\pi^3 c^2} \frac{\hbar \omega}
{\exp(\hbar \omega/k_{\rm B}T) -1} ,
\end{equation}
where $c$ is the speed of light. The black-body limit can be obtained by assuming that the absorption efficiencies 
$Q_{\rm TM, TE}(\omega)=1$ for all frequencies. In this case, equation~\ref{eq-Q} reduces to the Stefan-Boltzmann law 
\cite{Incropera2017}: $P_{\rm BB} = \sigma A F_{12}(T^4_1-T^4_2)$, where $\sigma = 5.67 \times 10^{-8}$ W/(m$^2$K$^4$).
It is worth mentioning that we have verified the validity of eq~\ref{eq-Q} to calculate the FFRHT between 2D materials by 
comparing its results with numerically exact simulations within the framework of fluctuational electrodynamics
(Supporting Information, Figure S1). 

According to eq~\ref{eq-Q}, $Q_{\rm TM, TE}(\omega)$ are required to calculate the FFRHT between two graphene sheets. 
Since our system is one-atom-thick in the $y$-direction (Figure~\ref{fig-setup}a), a wave impinging with the electric
field pointing in the $y$-direction does not generate any current on that direction. Hence, the absorption cross section
of TM plane waves vanishes ($Q_{\rm TM}(\omega)=0$) and only $Q_{\rm TE}(\omega)$ contributes to the FFRHT.
Note that free-space propagating waves cannot couple efficiently to surface plasmons in graphene, 
which lie far outside the light line, due to the large mismatch in in-plane momentum.
We have calculated this efficiency using COMSOL MULTIPHYSICS, where we
have modeled our system as a 3D parallelepiped with an effective dielectric constant \cite{Nikitin2013} (see Supporting
Information). In accordance with experimental evidence, we have taken $\delta_{\rm graphene}=0.37$ nm
for the thickness of a graphene monolayer\cite{Koh2011}. Figure~\ref{fig-setup}c 
shows $Q_{\rm TE}(\omega)$ as a function of the radiation frequency $\omega$ (solid lines) for different chemical 
potentials and for a scattering energy $\mathcal{E} _{\rm s}=10^{-4}$ eV. Notice that the absorption cross section is 
much larger than the geometrical one in the infrared frequency range, which shows that graphene is a very efficient 
broadband infrared absorber, even when the incident vector of the plane wave is parallel to the graphene sheet
(see inset in Figure~\ref{fig-setup}c). Notice also that $Q_{\rm TE}(\omega)$ increases for decreasing frequency, 
which is due to the increase of losses in the system (see Figure~\ref{fig-setup}b). 
 
In order to get further insight into the remarkable radiation absorption of a graphene flake, we have derived an 
analytical expression for $Q_{\rm TE}(\omega)$ in eq 3 as follows. The radiation absorption can be understood as a two-step 
process. First, a TE plane wave impinges in the graphene flake (see inset in Figure~\ref{fig-setup}c) 
and couples to the guiding modes of our system. Second, 
these modes propagate along the $z$-direction, while being progressively absorbed by the graphene flake. Taking 
both processes into account (Supporting Information, Figure S2), the frequency-dependent absorption efficiency 
can be expressed as
\begin{equation}
Q_{\rm TE}^{\rm an}(\omega) = \frac{ 1-e^{-2 {\rm Im} \{k_z\}L_z} } {\delta \, {\rm Im}\{k_{y,{\rm v}}\}}.
\label{eq-anal}
\end{equation}
Here, $k_z$ corresponds to the $z$-component of the graphene mode wave vector and $k_{y,{\rm v}}$ represents the 
$y$-component of the same wave vector in vacuum.  
In eq~\ref{eq-anal}, the factor $1/{\rm Im}\{k_{y,{\rm v}}\}$ is related to the coupling between the plane wave 
and the EM mode of the system, while the numerator $(1-e^{-2 {\rm Im} \{k_z\}L_z} )$ accounts for the absorption 
of the guiding wave along the graphene sheet. We have computed the dispersion relation
of the leaky guided modes of our system by using standard 
dielectric waveguide theory \cite{Marcuse1991} (see Supporting Information). In Figure~\ref{fig-setup}c
we show the analytical results for the absorption efficiency $Q_{\rm TE}^{\rm an}(\omega)$ (dashed lines) and,
as one can see, there is an excellent agreement with the exact numerical simulations. This agreement allows us to
conclude that the extraordinary absorption efficiency of a graphene flake in this configuration is due to the fact
that it behaves as a lossy waveguide that absorbs the radiation via the excitation of guided TE modes. In particular, 
because of the low impedance mismatch, the incident radiation is efficiently coupled into guided modes and is eventually 
absorbed. 

\begin{figure}[t]
\centering\includegraphics[width=16cm]{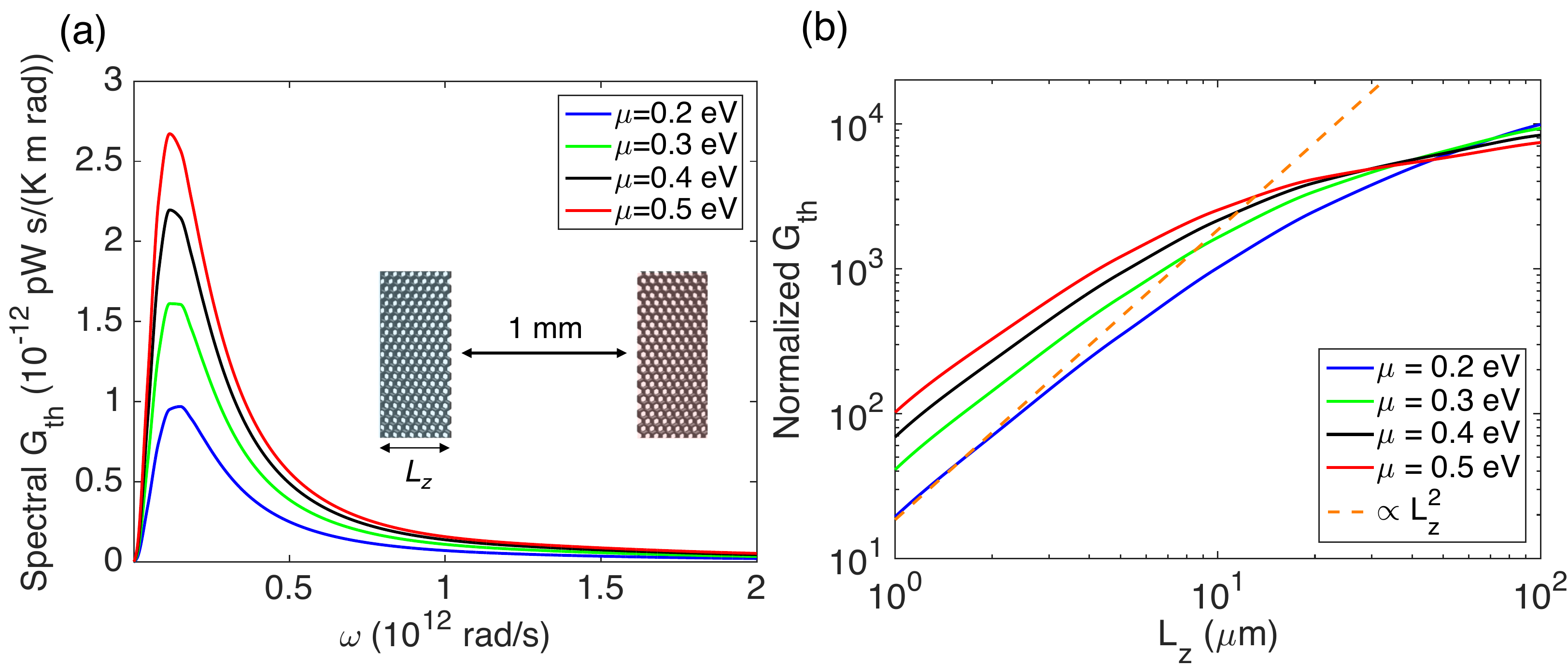}
\caption{(a) Spectral thermal conductance as a function of the radiation frequency for a system composed of two 
graphene flakes of length $L_z=10$ ${\rm \mu m}$, $\mathcal{E} _{\rm s}=10^{-4}$ eV, and a gap $d=1$ mm (see inset).
The temperature is 300 K. (b) The total thermal conductance $G_{\rm th}$, normalized by the blackbody results, for 
the same system as in panel (a) and plotted as a function of $L_z$ for different chemical potentials. The dashed 
orange line is proportional to $L_z^2$. Let us stress that these normalized results do not depend on the gap as long 
as $d$ is much larger than the thermal wavelength.}
\label{fig-normG}
\end{figure}

Once $Q_{\rm TE}(\omega)$ is known, we can use eq~\ref{eq-Q} to calculate the FFRHT between 
two graphene flakes in the coplanar configuration (see Figure~\ref{fig-setup}a). We shall characterize the FFRHT in 
terms of the room-temperature linear heat conductance per unit of length, $G_{\rm th}=P/(L_{x} \Delta T) $,
in the limit $\Delta T=(T_2 - T_1) \to 0$. Figure~\ref{fig-normG}a 
shows the spectral $G_{\rm th}$, i.e., the conductance per unit of frequency, for two graphene sheets of length 
$L_z=10$ ${\rm \mu m}$, $\mathcal{E} _{\rm s}=10^{-4}$ eV, and a gap $d=1$ mm. It can be observed that the system exhibits 
a broad-band FFRHT spectrum, similar to the FFRHT between metals \cite{Polder1971}. The conductance peak appears at  
$\omega=9\times10^{10}$ rad/s for all chemical potentials. This maximum originates from the convolution of the  
Planck's distribution function ($I_{\rm BB}$) and $Q_{\rm TE}(\omega)$, see eq~\ref{eq-Q}. Notice, however, that
the magnitude of $G_{\rm th}$ does increase with $\mu$ and can be tuned by a factor of 2.5 between $\mu= 0.2$ eV 
and $\mu = 0.5$ eV. 

Let us turn now to the analysis of the total thermal conductance and its comparison with the predictions of 
Planck's law for blackbodies. Figure~\ref{fig-normG}b shows $G_{\rm th}$ normalized by the corresponding 
blackbody result ($G_{\rm BB} = 4\sigma \delta F_{12}T^3$) as a function of the flake's length $L_z$. As it can be observed,
the power exchanged by the two graphene flakes overcomes Planck's results by up to 4 orders of magnitude for a length 
of 100 $\mu$m. The reason for this huge enhancement can be understood with the help of Figure~\ref{fig-setup}c, where 
it is shown that the absorption efficiency of a graphene sheet reaches values much larger than 1 for a broad range of 
infrared frequencies, accesible at room temperature. Besides, the normalized $G_{\rm th}$ increases with 
$L_{z}$ simply because the absorption and emission of radiation in the graphene flakes increases with this length. It can be also seen in 
Figure~\ref{fig-normG}b that for small lengths, the normalized $G_{\rm th}$ is proportional to $L_{z}^{2}$, which can
be understood as follows. The efficiency $Q_{\rm TE}(\omega)$ is proportional to $(1-e^{-2 {\rm Im} \{k_z\}L_z} )$, 
according to eq~\ref{eq-anal}. In the limit in which ${\rm Im} \{k_z\}L_z \ll 1$, $Q_{\rm TE}(\omega)$ is simply proportional 
to $L_z$. Thus, from eq~\ref{eq-Q}, it is obvious that $G_{\rm th} \propto L_{z}^2$ for short graphene flakes, as it 
is verified in Figure~\ref{fig-normG}b.

\begin{figure}[t]
\centering\includegraphics[width=13cm]{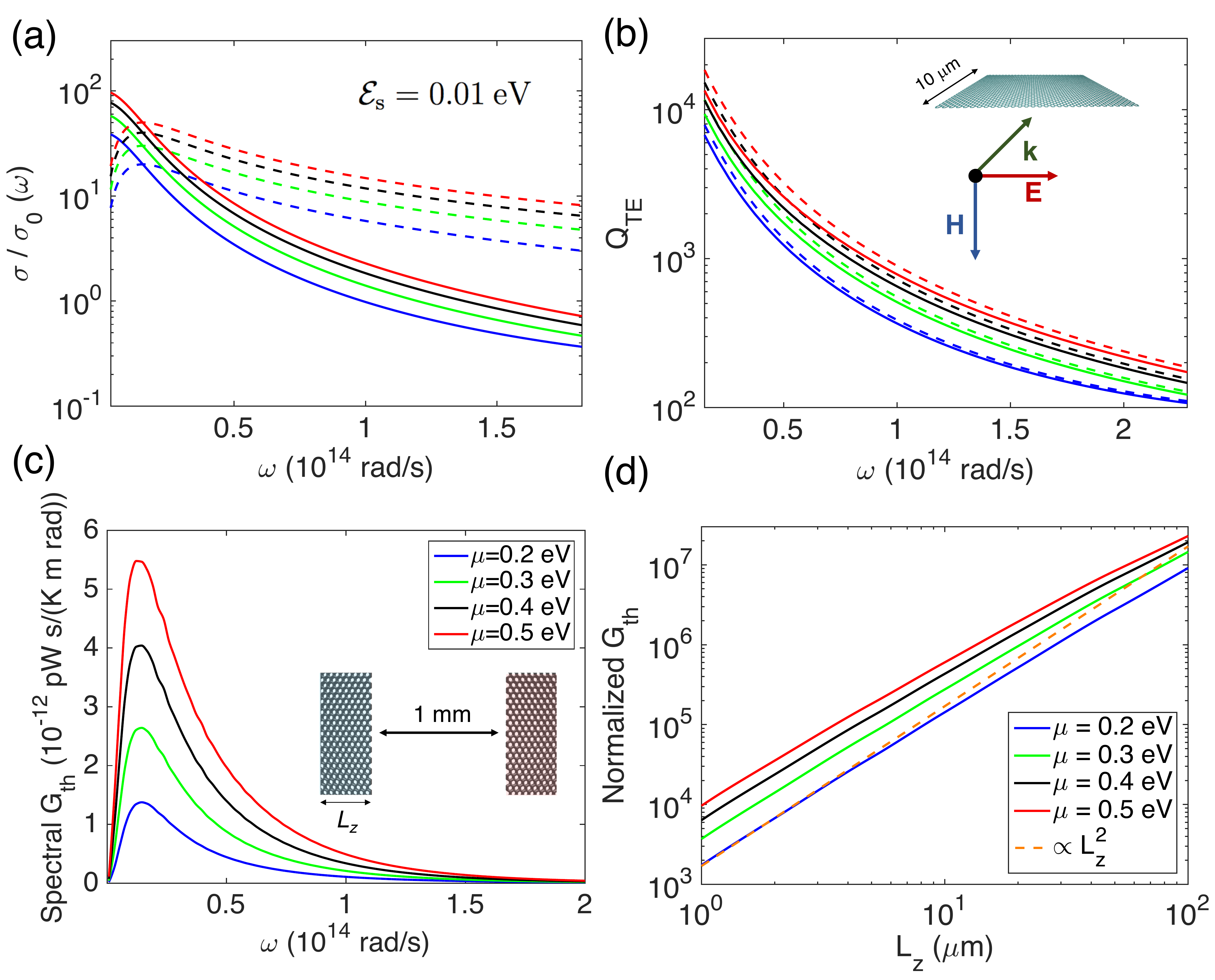}
\caption{ (a) Real (solid lines) and imaginary (dashed lines) part of the normalized conductivity of graphene for 
$T=300$ K and $\mathcal{E} _{\rm s}=0.01$ eV, for different chemical potentials ($\mu$). (b) Frequency-dependent 
absorption efficiency for a plane wave with transverse electric polarization ($Q_{\rm TE}(\omega)$) and normal 
incidence into a graphene sheet with length $L_z = 10 \,\, \mu$m and infinite width (see inset), for different values of 
$\mu$. The solid lines correspond to the exact numerical results, while the dashed lines were obtained with 
eq~\ref{eq-anal}. (c) Spectral $G_{\rm th}$ as a function of $\omega$ for a system composed of two graphene 
flakes of length $L_z=10$ ${\rm \mu m}$, $\mathcal{E} _{\rm s}=0.01$ eV, and a gap $d=1$ mm (see inset). 
(d) $G_{\rm th}$, normalized by the blackbody results, for the same system and plotted as a function of $L_z$ 
for different chemical potentials. The dashed orange line is proportional to $L_z^2$.}
\label{fig-realgraph}
\end{figure}

We have shown that the FFRHT between graphene sheets can overcome the Planckian limit by more
than 4 orders of magnitude. However, our analysis also suggests that the thermal conductance could be further enhanced
by increasing the intrinsic losses in the graphene sheets. To test this idea, we have calculated the FFRHT 
for these graphene sheets assuming a larger value for the scattering energy $\mathcal{E}_{s}$.  
Figure~\ref{fig-realgraph}a shows the normalized 2D conductivity of graphene for $\mathcal{E}_{\rm s}=0.01$ eV,
i.e., two orders of magnitude larger than in the examples above. The corresponding results for the absorption
efficiency $Q_{\rm TE}(\omega)$ are displayed in Figure~\ref{fig-realgraph}b.
The absorption cross section is again orders of magnitude larger than the geometrical one and, more importantly,
it is also higher than in the previous case. Figure~\ref{fig-realgraph}c shows the spectral $G_{\rm th}$ 
for $L_z=10$ ${\rm \mu m}$, $\mathcal{E} _{\rm s}=0.01$ eV, and a gap
$d=1$ mm. In this case the maximum of the spectral $G_{\rm th}$ is strongly blueshifted ($\omega=1.3\times10^{13}$ rad/s), 
and the relevant frequencies for the FFRHT are also higher. The reason for this blueshift is that $Q_{\rm TE}(\omega)$
adopts larger values at frequencies which have a better overlap with Planck distribution function at
room temperature. As a consequence, the total thermal conductance $G_{\rm th}$ is much higher in this case, 
as we illustrate in Figure~\ref{fig-realgraph}d. Notice that in this case the Planckian limit can be overcome 
by more than 7 orders of magnitude. Thus, we see here that the graphene with a high density of impurities (i.e., 
with low mobility), which is normally dismissed for optoelectronic and plasmonic applications, is more efficient 
regarding thermal emission and absorption.

For the sake of comparison, we have also analyzed the FFRHT between two graphene sheets of the 
same dimensions as those of Figure~\ref{fig-realgraph}c ($\mu=0.3$ eV) now parallel to each other 
and separated by a distance $d$ along the normal direction. In that case, the geometrical
cross section is 27000 times larger than in the 
coplanar configuration and Planck's law would thus predict 27000$^2$ higher heat transfer efficiency 
than in the coplanar case. However, the FFRHT between the graphene sheets in this case is only 4 times 
larger and it does not exhibit an enhancement over Planck's law.
Indeed, the ratio with the blackbody results is $1.5\times10^{-3}$. This confirms that
the FFRHT between coplanar sheets is truly remarkable and that its absolute value is comparable with other setups that have
a much higher geometrical cross section. Moreover, we have performed additional simulations to verify if such FFRHT
could be measured in a realistic experimental setup. 
We have calculated the FFRHT between two graphene sheets with $L_x=20$ $\mu$m, $L_z=60$ $\mu$m, 
$\mu=0.5$ eV, $\mathcal{E} _{s}=0.01$ eV and separated by a gap of 20 $\mu$m, where the thermal radiation is 
already dominated by the far-field contribution \cite{Song2015}. The dimensions chosen for both the graphene 
sheets and the gap are within reach of state-of-the-art calorimetric techniques\cite{Sadat2013,Zheng2013}. In 
order to compute the FFRHT, we have made use of the code SCUFF-EM, which implements a fluctuating-surface-current 
approach to the radiative heat transfer problem and provides numerically exact results within the framework of fluctuational 
electrodynamics\cite{Rodriguez2013,Reid2015}. The room-temperature linear heat conductance between the 
flakes is in this case 1.62 pW/K, which is within the sensitivity of existent calorimetric techniques\cite{Sadat2013,Zheng2013}.

\begin{figure}[t!]
\centering\includegraphics[width=13cm]{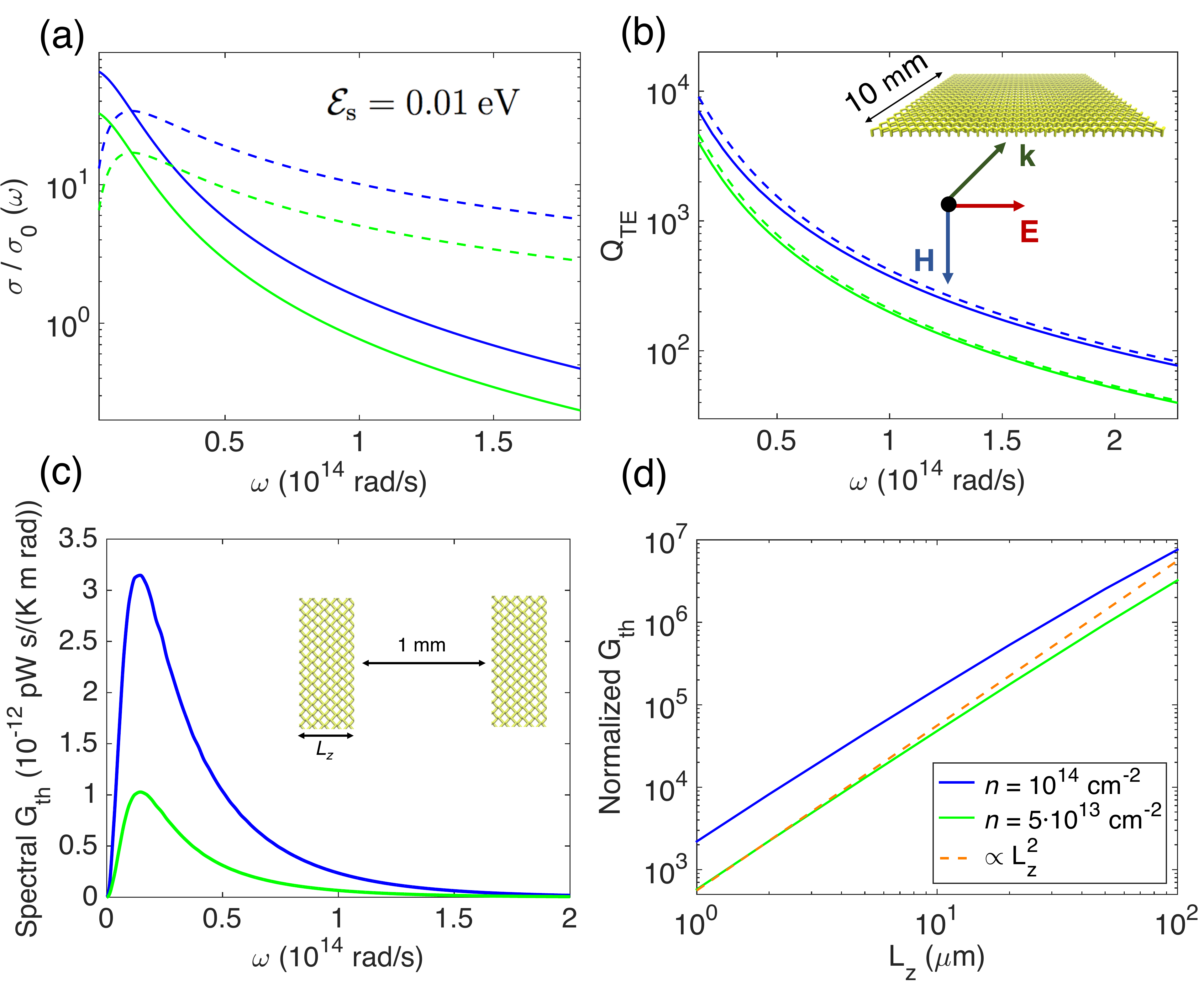}
\caption{(a) Real (solid lines) and imaginary (dashed lines) part of the normalized conductivity of single-layer black phosphorus for 
$T=300$ K and $\mathcal{E} _{\rm s}=0.01$ eV, for two different dopings ($n$). (b) Frequency-dependent 
absorption efficiency for a plane wave with transverse electric polarization ($Q_{\rm TE}(\omega)$) and normal incidence 
into a SLBP sheet with length $L_z=10$ $\mu$m and infinite width (see inset). The solid lines correspond to
the exact numerical results, while the dashed lines were obtained with eq~\ref{eq-anal}. (c) Spectral $G_{\rm th}$ as a function of 
$\omega$ for a system composed of two SLBP flakes of length $L_z=10$ ${\rm \mu m}$ separated by a gap $d=1$ mm (see inset). 
(d) Thermal conductance $G_{\rm th}$, normalized by the blackbody results, for the same system as in panel (c) and plotted 
as a function of $L_z$, for two different dopings. The dashed orange line is proportional to $L_z^2$.}
\label{fig-BP}
\end{figure}

At this point one may wonder whether the dramatic violation of Planck's law discussed above for the case of graphene
may also occur in other 2D materials. To show that this is actually the case, we now turn to analyze the case of 
single-layer black phosphorus (SLBP). We have computed the FFRHT between two coplanar SLBP sheets (see 
Figure~\ref{fig-setup}a). The distinctive steps of the atomic structure of SLBP are in our case placed along the 
$z$-direction. We have modeled the dielectric properties of a black phosphorus monolayer in an analogous way to graphene 
and its 2D conductivity has been taken from previous studies \cite{Low2014}. Both the real (solid line) and the imaginary 
(dashed line) part of the conductivity of  SLBP along the $x$-direction are plotted in Figure~\ref{fig-BP}a for $T=300$ K, $\mathcal{E} _{s}=0.01$ eV,
and two different electron dopings $n=5 \times 10^{13}$ cm$^{-2}$ and $n=10^{14}$ cm$^{-2}$. 
The parameters chosen represent realistic SLBP samples\cite{Saito2015}. Figure~\ref{fig-BP}b shows $Q_{\rm TE}(\omega)$ calculated 
numerically (solid lines) with COMSOL MULTIPHYSICS (see Supporting Information) for a SLBP sheet with $L_z=10$ $\mu$m 
(see inset of Figure~\ref{fig-BP}b) and both doping values. The SLBP absorption efficiency exhibits 
very similar characteristics to those of low-quality graphene, as both of them have similar dielectric functions for infrared 
frequencies. Moreover, $Q_{\rm TE}^{\rm an}(\omega)$ (dashed lines) shows again an excellent agreement with the exact numerical 
simulations. As for graphene, we have used the results for $Q_{\rm TE}(\omega)$ in combination with eq~\ref{eq-Q} to 
describe the FFRHT. The spectral conductance of black phosphorus monolayers separated by 1 mm is
plotted in Figure~\ref{fig-BP}c, while the normalized total thermal conductance as a function of the length $L_z$ is
shown in Figure~\ref{fig-BP}d. Notice that in this case the FFRHT can be larger than the corresponding result
calculated from Planck's law by almost 7 orders of magnitude, showing that this enhancement is not exclusive of graphene, 
but can also occur in other 2D materials such as SLBP.

In summary, we have presented a theoretical analysis of the FFRHT between 2D materials, graphene and single-layer 
black phosphorus, in a coplanar configuration. We have shown that the relevant absorption cross section of flakes of these 
materials can be orders of magnitude larger than their atomic-sized geometrical cross section. We have also shown 
that this extraordinary absorption efficiency makes the FFRHT between flakes of these 
materials more than 7 orders of magnitude larger than the limit set by Planck's law, which constitutes the ultimate 
violation of this law in the far-field regime. Finally, we have shown that the novel mechanism responsible for this FFRHT 
involves the propagation properties of TE-polarized guiding modes in these materials, modes that are usually irrelevant in 
the context of plasmonic or optoelectronic applications.

\begin{acknowledgement}

This work has been financially supported by the Spanish MINECO (FIS2015-64951-R, MAT2014-53432-C5-5-R, 
FIS2017-84057-P), the Comunidad de Madrid (S2013/MIT-2740), the European Union Seventh Framework 
Programme (FP7-PEOPLE-2013-CIG-630996), and the European Research Council 
(ERC-2011-AdG-290981 and ERC-2016-STG-714870). V.F.-H.\ acknowledges support from `la Caixa' Foundation. 
V.F-H.\ and J.C.C.\ (Mercator Fellow) thank the DFG and SFB767 for sponsoring their stay at the University of Konstanz.\\
\end{acknowledgement}

\begin{suppinfo}
Proof of the validity of eq~3 to calculate the FFRHT between 2D materials, details on the 
modeling of 2D materials as 3D parallelepipeds, and the derivation of eq~5.
\end{suppinfo}

\end{document}